\documentclass{aastex}

\shorttitle{Nature of CFC97 Cen 05}

\shortauthors{Bouchard, Da Costa \& Jerjen}

\begin{document}

\title{Clarification of the Nature of the Galaxy CFC97 Cen 05\footnote{Based in part on 
observations made with the NASA/ESA Hubble Space Telescope, obtained from 
the data archive at the Space Telescope Science Institute.  STScI is operated 
by the Association of Universities for Research in Astronomy, Inc.\ under 
NASA contract NAS 5-26555.}}

\author{A. Bouchard, G. S. Da Costa and  H. Jerjen}
\affil{Research School of Astronomy \& Astrophysics, Institute of 
Advanced Studies, Australian National University, Mt Stromlo Observatory,
Cotter Road, Weston Creek, ACT 2611, Australia}
\email{bouchard, gdc, jerjen @mso.anu.edu.au}

\begin{abstract}
The galaxy CFC97 Cen 05 has in the past been considered an
H{\small I}-rich dwarf galaxy in the nearby Centaurus~A group.  We have
used Australia Telescope Compact Array observations to show
that the H{\small I} associated with CFC97 Cen 05 by \citet{CO97} 
is most likely a Galactic High Velocity Cloud that
is centered $\sim$17$\arcmin$ from the optical image of the
galaxy. At the optical location of the galaxy, which is not that 
tabulated by \citet{BA99}, there is no indication of the 
presence of H{\small I} for velocities less than the upper limit
of the H{\small I}PASS survey at $\sim$12,500 kms$^{-1}$.  In addition,
WFPC2 images of CFC97 Cen 05 obtained from the {\it  HST} Archive
reveal that this galaxy is in fact not a dwarf at all, but rather is a 
distant background spiral not associated with the Centaurus~A group.
\end{abstract}


\section{Introduction}

The current paradigm for the formation of structure in the Universe
($\Lambda$CDM) envisages that relatively low-mass 
\citep[$\sim10^7\/ M_\sun$, e.g.][]{LM02}
dark matter halos form first from density fluctuations in the early 
Universe.  These small-scale clumps of dark matter then merge in a 
hierarchical process to form larger structures.  This model has revived
interest in dwarf galaxies since such systems may well represent the closest
present-day analogues to the products of the evolution of the initial dark 
matter halos.  Dwarf galaxies are ubiquitous, being found in environments as 
diverse as rich clusters through to the general field.  In many instances,
the study of dwarf galaxies and the influence of environment on their
evolution is based on observations of dwarfs in relatively nearby
groups.  The proximity of such systems allows the investigation of 
the properties of the dwarfs over a substantial range in luminosity, 
including systems approaching the absolute magnitude of all but the 
faintest of the Local Group dwarf population.  

Frequently, the initial classification of a system as potential 
dwarf galaxy member of a group is done via visual inspection of Sky Survey
films \citep[e.g.][]{CO97, HB00}. 
Such classification as a dwarf candidate is necessarily a subjective
process: morphology of the image, surface brightness, extent of resolvable 
structure, etc, all play a role in the categorization process.  Subsequent
follow-up analyses necessarily require the census of dwarfs in a group to 
be as complete as possible.  Consequently, confirmation of group
membership needs to be carried forward for the dwarf candidates.  This
can come about via measurement of a radial velocity consistent
with group membership, either through optical spectra or H{\small I} 
observations, or via a determination of a distance compatible with group
membership obtained either directly from, for example, the tip of the 
red giant branch, or through surface brightness fluctuation techniques
\citep[e.g.][]{HJ00}.

In this paper we present a discussion of the galaxy CFC97 Cen 05.  This
galaxy was discovered and classified visually by \citet{CO97}
as one of a number of potential dwarf irregular members of the Centaurus~A 
group.  This
group lies at a distance of $\sim$3.5 Mpc, and as the name implies, has the 
unusual E galaxy Cen~A (NGC~5128) as the dominant group member.  
\citet{CO97} sought to confirm group membership of their candidates 
through the measurement of radial velocities obtained with either the Parkes 
radio-telescope at 21 cm, and/or with the Siding Spring Observatory 2.3 m 
telescope at H$\alpha$\@.  
Although the heliocentric velocity for CFC97 Cen 05 given by \citet{CO97},
+122 kms$^{-1}$ obtained from H{\small I} observations, is more than 
200 kms$^{-1}$ lower than the next lowest velocity of the candidates 
with H{\small I} detections (ESO~383-G087)\footnote{Cen 05 is
not plotted in the lower panel of Fig.\ 5 of \citet{CO97}, as its V$_{LG}$,
calculated using the definition employed by \citet{CO97}, is
--166 kms$^{-1}$.}, \citet{CO97} classified CFC97~Cen 05 as a confirmed dwarf 
member of the Cen~A group, and listed the total H{\small I} content as 
1.6 $\times$ 10$^{7}\/M_{\sun}$.  Assuming membership in the Cen~A group
and using the photometry of \citet{SC95} corrected for interstellar
extinction but not absorption internal to the galaxy, the absolute blue
magnitude of CFC97 Cen 05 is M$_{B}$ $\approx$ --12.0, yielding 
M$_{H\small I}$/L$_{B}$ $\approx$ 1.7 in solar units, a value typical
of most dwarf irregulars \citep{BW04}.

CFC97 Cen 05 is also mentioned in the blind H{\small I} survey of the
Cen A group carried out by \citet{BA99} with the Parkes telescope
and the multibeam focal plane array system \citep{BD01}.  \citet{BA99} 
listed an
H{\small I} detection with a heliocentric velocity of 130 kms$^{-1}$
that corresponded to an  H{\small I} mass of 1.7 $\times$ 10$^{7}\/M_{\sun}$
assuming membership in the Cen~A group.  These values agreed with those 
of \citet{CO97} for CFC97 Cen 05 and consequently, \citet{BA99} assigned 
their H{\small I} detection to this object.  However, \citet{BA99} apparently
failed to notice that the position for the object they claimed as the 
detection of H{\small I} in CFC97 Cen 05 is in fact $\sim$17$\arcmin$ from 
the location of the optical image of the galaxy.  

As part of a larger project investigating the influence of environment on
the evolution of dwarf galaxies, we are carrying out a study of the
distribution of the H{\small I} in a number of dwarf galaxies in 
nearby groups, including the Cen~A group.  Local Group observations
show that in low luminosity dwarfs, the gas can be centered on the optical 
component, offset from
the center of the optical component, or perhaps even completely detached
\citep[e.g.][]{YK97,SG99,RS02}. 
While the Parkes telescope is adequate in most cases for detecting the 
presence of H{\small I}, given the small angular size of the target dwarfs,
and the large beam size of that telescope ($\sim$15$\arcmin$), 
higher spatial resolution observations are required to investigate the 
location of the neutral gas relative to the optical component.  Consequently,
we have employed the Australia Telescope Compact Array (ATCA) for our 
study.
One of our targets was CFC97 Cen 05, which lies 
$\sim$7$\arcmin$ ($\sim$7 kpc in projection) from the luminous spiral 
NGC~4945 that has a Seyfert~2 nucleus.  Our ATCA
observations and subsequent follow-up using HST Archive data have revealed 
that not only is the
H{\small I} detected by \citet{CO97} and \citet{BA99} not associated
with CFC97 Cen 05, it is instead most likely a foreground Galactic High 
Velocity Cloud (HVC), but also the galaxy itself is background spiral and 
not a dwarf in the Cen~A group.

The remainder of the paper is arranged as follows.  In the following section
we present and discuss our ATCA observations of CFC97 Cen 05, while in
Sect.\ 3 we present an optical image of the galaxy obtained from HST
Archive WFPC2 images.  Our results are summarized in the final section.

\section{Australia Telescope Compact Array Observations}

CFC97 Cen 05 was observed with the ATCA on 24 Feb 2003.  
A full 12 hour
integration was carried out using the 750D array configuration. 
The FULL\_4\_1024-128 correlator configuration was employed with a 4 MHz 
bandpass centered on 1420 MHz, yielding velocity coverage from 
--315 to +528 kms$^{-1}$ divided into 1024 channels, each of width 0.82 
kms$^{-1}$.  The primary beam has a FWHM of 33.6$\arcmin$ and it was
centered at the CFC97 Cen 05 position as listed in \citet{BA99}.
The data were reduced using the MIRIAD package. The resulting cube has 
a circular beam 2$\arcmin$ in diameter and the velocity axis was 
binned to a channel spacing of 4.1 kms$^{-1}$.

\begin{figure}
\plotone{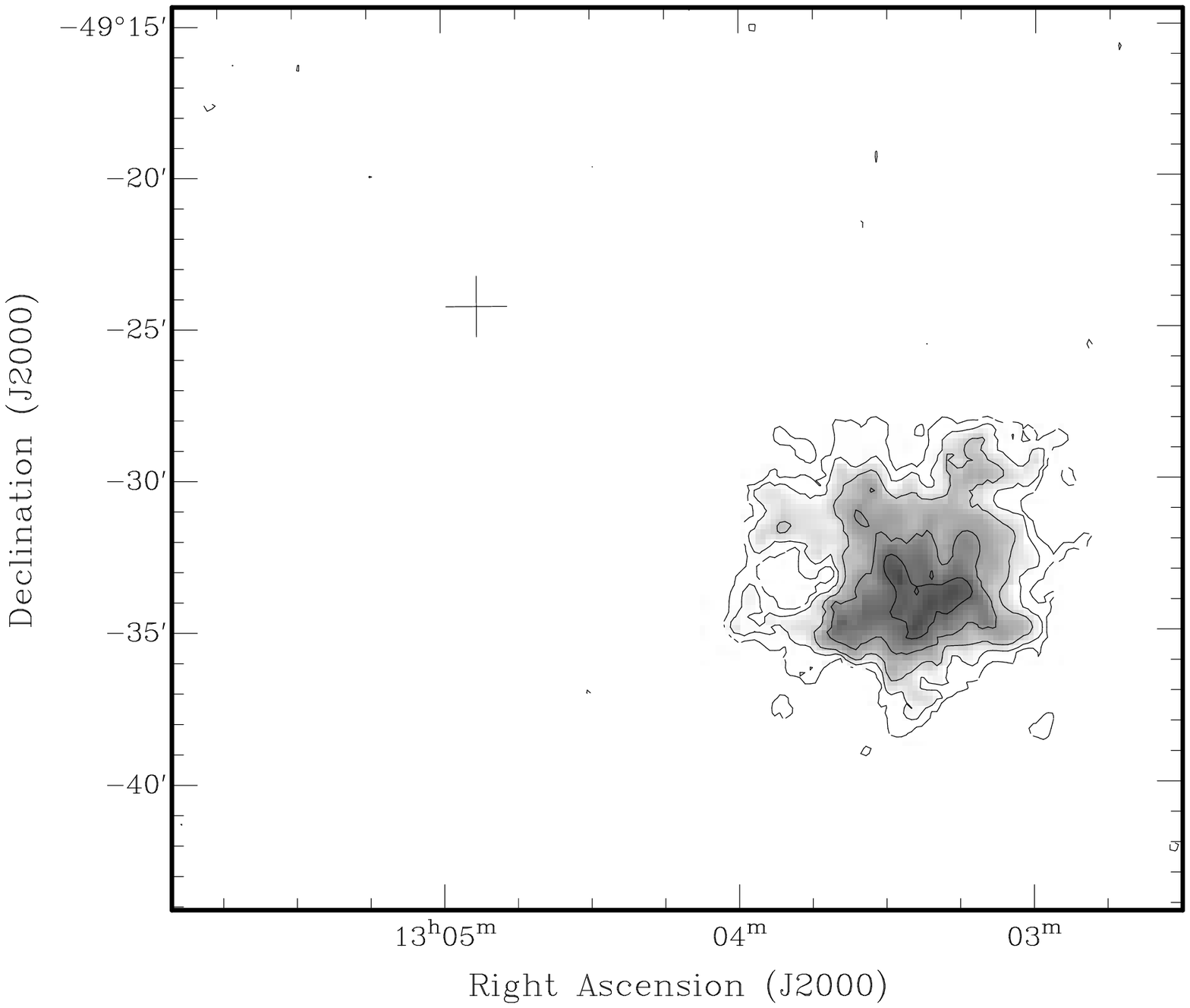}
\caption{An H{\small I} intensity map from the ATCA 
observations for the velocity range 109 to 142 kms$^{-1}$.  
The column density levels are 0.5, 1.0, 1.5, 2.0 and 2.5 $\times$ 10$^{19}$
cm$^{-2}$.  This gas cloud is that detected with the 
Parkes telescope
by \citet{CO97} and \citet{BA99}, and associated by them with CFC97 Cen 05.
However, the plus symbol to the North-East of the H{\small I} cloud marks 
the location of the optical image of CFC97 Cen 05.  The symbol is 
$2\arcmin\times2\arcmin$ in size and encompasses the optical extent
of the galaxy.  The $\sim$17 kpc
separation between the optical image of the galaxy and the 
H{\small I} cloud, if both are at the distance of the Cen~A group, makes 
the claimed association unlikely. \label{AB_hvc.eps} }
\end{figure}

Both \citet{CO97} and \citet{BA99} report the detection of H{\small I}
at a heliocentric velocity of approximately 125 kms$^{-1}$.  We show in
Fig.\ \ref{AB_hvc.eps} a map of the H{\small I} detected in our ATCA
observations for the velocity range of 109 to 142 kms$^{-1}$.  
The map 
shows a relatively diffuse cloud $\sim$10$\arcmin$ in diameter, whose
center is coincident with the position listed by \citet{BA99}
for their claimed detection of H{\small I} in CFC97 Cen 05.  
The cloud has a mean heliocentric velocity of 124.7 kms$^{-1}$ with a
velocity width W$_{20}$ of 36 kms$^{-1}$.  These values are very
similar to those given by \citet{CO97}, 122 and 53 kms$^{-1}$, and by
\citet{BA99}, 130 and 46 kms$^{-1}$, respectively.  The total flux
from the cloud is 4.5 Jy kms$^{-1}$, which is somewhat less than the
total fluxes given by \citet{CO97} and \citet{BA99}, 5.5 and 5.9 
Jy kms$^{-1}$, respectively.  Such a discrepancy is not unexpected
given our observations are with an interferometer rather than a 
single dish.

However, as shown in Fig.\ \ref{AB_hvc.eps}, the center of the H{\small I} 
cloud is $\sim$17$\arcmin$ from the optical position of the galaxy.
Assuming both the gas cloud and CFC97 Cen 05 are in the Cen~A group at
a distance of 3.5 Mpc, the diameter of the H{\small I} cloud would be
approximately 10 kpc and the projected separation from the optical galaxy 
would be of order 17 kpc.  Given that CFC97 Cen 05 has an optical
diameter of $\sim$1 kpc if it is in the Cen~A group, it would seem very
unlikely that such a large H{\small I} cloud is physically associated 
with CFC97 Cen 05, if both are at the distance of the Cen~A group.

We are left then with two distinct questions.  First, is the H{\small I}
cloud actually associated with the Cen~A group, and second, is there
any evidence for H{\small I} at, or near, the optical position of CFC97
Cen 05 that might be associated with that system?  As regards the first
question, we note that inspection of the Digital Sky Survey shows no
sign of any obvious optical counterpart at the location of the H{\small I} 
cloud, though interpretation of the DSS image is complicated by the fact 
that the bright star HD113314 ($V$ = 4.8) lies on the sky only 
1.8$\arcmin$ from the center of the
H{\small I} cloud.  Nevertheless, given the results of the H{\small I}PASS 
survey, which has revealed that isolated H{\small I} clouds beyond the Local 
Group without optical counterparts are extremely rare 
\citep{KB04}, it 
would seem rather unlikely that this cloud is associated with the Cen~A 
group.  Further, from our observations,
the velocity of the Cloud relative to the Local Group 
\citep[using the same definition as][]{CO97} is
--163 kms$^{-1}$, which is $\sim$240 kms$^{-1}$ 
less than the velocity of the Cen~A group member (ESO 383-G087) with
the lowest velocity relative to the Local Group.  As Fig.\ 5 of \citet{CO97}
indicates, the bulk of the galaxies of the Cen~A group have velocities
relative to the Local Group in the range of 75 to 400 kms$^{-1}$.  
Consequently, we believe that the
most likely interpretation of the H{\small I} cloud is that is a Galactic
HVC and thus not associated in anyway with the Cen~A group.  The cloud is 
not listed in the HVC catalog of \citet{MP02} though
this is not surprising, as with an area of $\sim$0.02 deg$^2$, it is
below the resolution limit of that survey.  The \citet{MP02} catalog, 
however, does list 8 HVCs within a 5 deg radius of the cloud.  These clouds
have  
V$_{LSR}$ velocities ranging from 90 to 210 kms$^{-1}$.  Given that
the V$_{LSR}$ of the H{\small I} cloud detected here is $\sim$120 kms$^{-1}$,
our interpretation of the cloud as a HVC is plausible.

\begin{figure}
\plotone{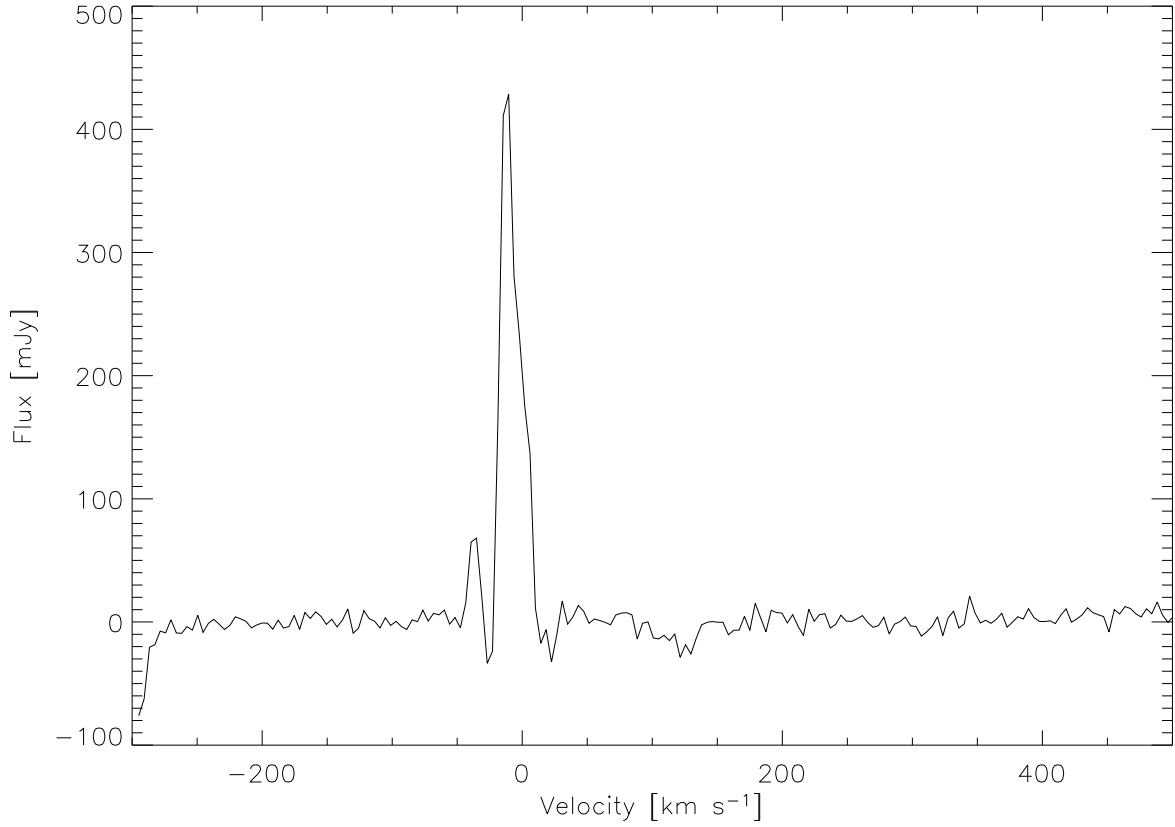}
\caption{A H{\small I} spectrum from the ATCA 
observations
for a $2\arcmin\times2\arcmin$ region centered on the optical image of 
CFC97 Cen 05. 
The RMS of this spectrum is 5.7mJy. 
Aside from the Galactic H{\small I} near zero velocity, there is no 
indication of any H{\small I} at this location within this velocity
interval. \label{AB_cen5spec.ps} }
\end{figure}

We turn now to the second question: whether there is any H{\small I} in the
vicinity of the optical image of CFC97 Cen 05.  In Fig.\ \ref{AB_cen5spec.ps}
we show the ATCA spectrum for a $2\arcmin\times2\arcmin$ region centered 
on the location
of the optical image of CFC97 Cen 05.  Aside from the Galactic H{\small I}
detected at velocities near zero, there is no indication of the presence
of any H{\small I} within the velocity range observed.  This spectrum has
an RMS of 5.7 mJy which converts to a 3$\sigma$ upper limit on the 
mass of any H{\small I} present of $4\times10^{3}$ $[D$/Mpc$]^2$ 
$[\Delta V/$kms$^{-1}]$ M$_{\odot}$, where $D$ is the distance and 
$\Delta V$ is the velocity width of the H{\small I}\@.  At the distance 
of the Cen~A group (3.5 Mpc) this limit is 7.4 $\times$ 10$^{5}$ M$_{\odot}$ 
for an assumed velocity width of 15 kms$^{-1}$.  

\begin{figure}
\plotone{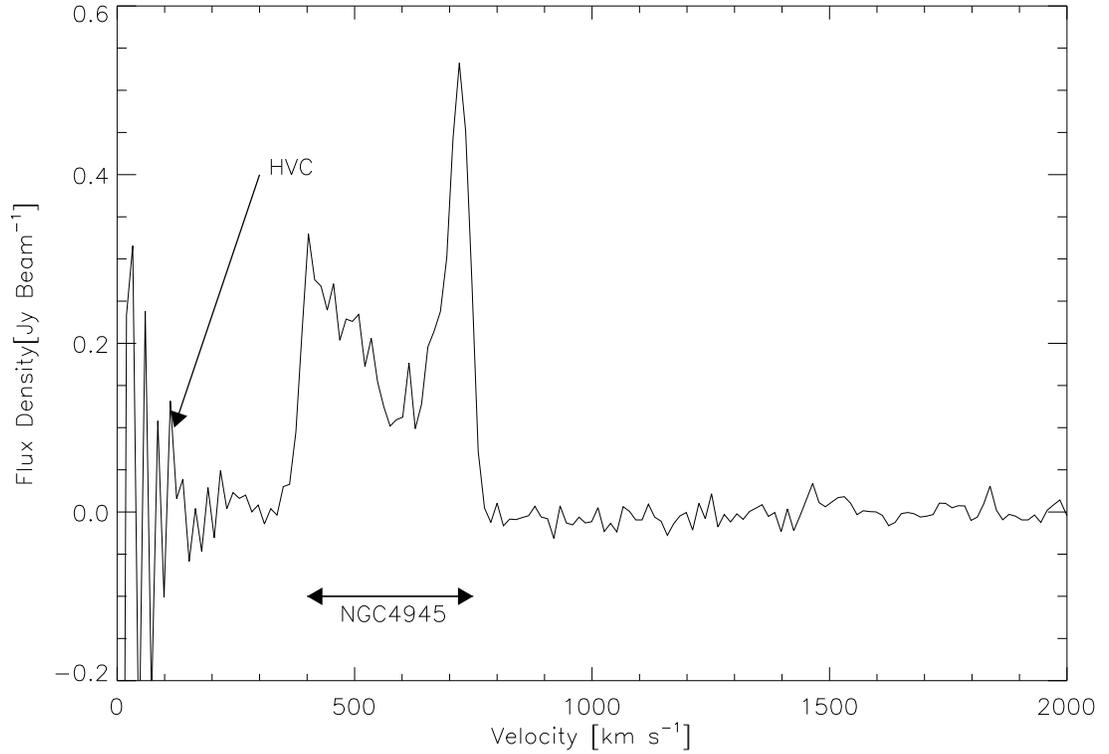}
\caption{The H{\small I}PASS spectrum at the optical
location of CFC97 Cen 05.  The lower spatial resolution of the 
H{\small I}PASS data results in contamination from the nearby spiral
galaxy NGC~4945.  H{\small I} from the HVC can also be identified at
a velocity of $\sim$120 kms$^{-1}$.  No other H{\small I} detections
are evident. \label{AB_HIPASS.ps} }
\end{figure}

It is of course conceiveable that there may be H{\small I} associated with 
CFC97 Cen 05 that lies at velocities exceeding the upper limit of our ATCA 
observations (+528 kms$^{-1}$).  To investigate this we show in 
Fig.\ \ref{AB_HIPASS.ps},
a H{\small I}PASS spectrum centered on the optical location of CFC97 Cen 05
for velocities between 0 and 2000 kms$^{-1}$.  Given that the spatial
resolution of the H{\small I}PASS data is $\sim$15$\arcmin$, it is
not surprising that this spectrum also shows H{\small I} from  
the spiral galaxy NGC~4945 that lies nearby on the sky. H{\small I} from 
the HVC can also
be identified at a velocity of $\sim$120 kms$^{-1}$.  However, there no are
indications of any other H{\small I} detections in this spectrum.  Further,
this result applies out to the upper velocity
limit of the H{\small I}PASS data at $\sim$12,500 kms$^{-1}$.  The
typical RMS in H{\small I}PASS data is 13 mJy \citep{BD01} leading
to a 3$\sigma$ detection limit of $\sim$10$^{4}$ $[D$/Mpc$]^2$ 
$[\Delta V/$kms$^{-1}]$ M$_{\odot}$.  At the distance of the Cen~A group
and for a velocity dispersion of 15 kms$^{-1}$, this corresponds to a
H{\small I} mass limit of 1.7$\times10^{6}$ M$_{\odot}$.  While further ATCA 
observations
are required to definitely rule out the possibility of H{\small I} at the
location of CFC97 Cen 05 which falls within the velocity range obscured by
NGC~4945 in Fig.\ \ref{AB_HIPASS.ps}, it seems unlikely that there is
any H{\small I} at this location, at least for velocities less than
$\sim$12,500 kms$^{-1}$.

\section{HST Archive Data}

Our study of the dwarf galaxies of the Cen~A group makes use, wherever
possible, of HST Archive data to, for example, determine distances
from the $I$ magnitude of the tip of the red giant branch 
\citep[cf.][]{KS02}.  Consequently, we investigated whether the 
HST Archive contained any data for CFC97 Cen 05.  Fortunately,
this galaxy was included in the field-of-view of a series of images taken 
with the WFPC2 camera on 25 March 1998 as a pure parallel observation 
(program 7909).  The data comprise four exposures taken through the 
$F606W$ 
filter (wide-$V$) with exposure times of 260, 500, 500 and 700 seconds, 
respectively.  CFC97 Cen 05 is located on the WF2 CCD in 
each image.

\begin{figure}
\plotone{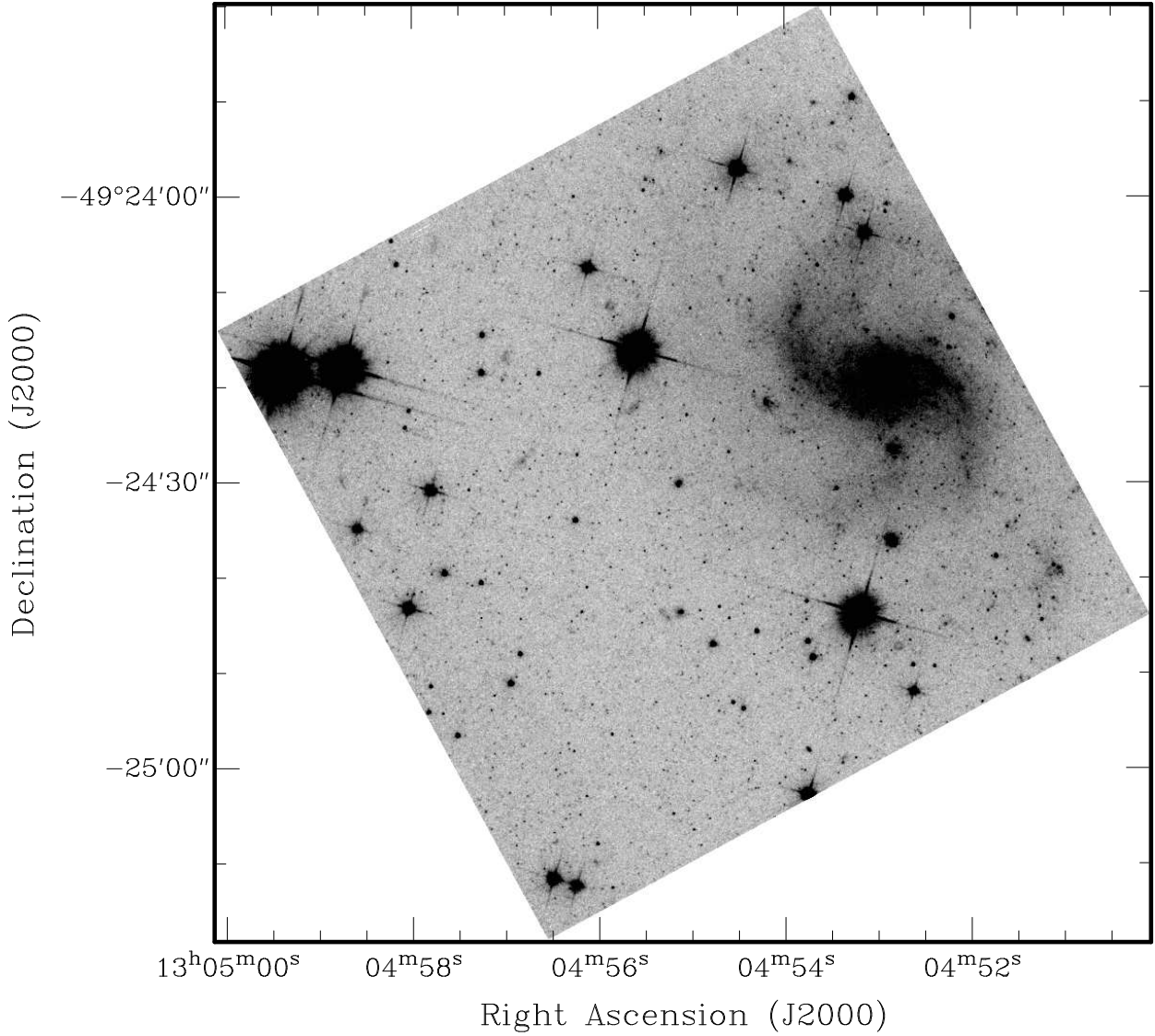}
\caption{An image of the galaxy CFC97 Cen 05 generated
from three HST WFPC2 images in the $F606W$ (wide-$V$) filter.  The
$1.2\arcmin\times1.2\arcmin$ field of the WF2 CCD is shown.   The galaxy 
is evidently a background spiral rather than a dwarf member of the Cen~A 
group.  \label{wf2pic} }
\end{figure}

The processed data frames for the three longer exposure images were
downloaded from the Archive and combined using standard techniques
\citep[cf.][]{DA02}.  The resulting image of CFC97 Cen 05 
is shown in Fig.\ \ref{wf2pic}.  It is 
immediately evident from this image that CFC97 Cen 05 is not a dwarf
galaxy at all; rather it is an almost face-on spiral galaxy
that is presumably located well beyond the Cen~A group.  On this image the 
disk of the galaxy is traceable out to a diameter of $\sim$60$\arcsec$ and
numerous likely H{\small II} regions are evident.
An optical spectrum should then be able to readily detect H$\alpha$ emission 
from this galaxy and thus determine its redshift.  At the present time we can 
only note that the lack of any H{\small I} detection at the location of this 
galaxy in the H{\small I}PASS survey suggests that the redshift
exceeds $\sim$12,500 kms$^{-1}$, the limit of that survey.

\section{Conclusions}

We have used H{\small I} observations with the Australia Telescope
Compact Array and HST Archive data to clarify the nature of the
galaxy CFC97 Cen 05, reputedly a relatively gas-rich dwarf in the
nearby Cen~A group.  The H{\small I} observations reveal that the
gas previously associated with this galaxy is instead a Galactic
High Velocity Cloud centered some $\sim$17$\arcmin$ from the optical
image of CFC97 Cen 05.  No gas is detected at the optical location 
of the galaxy for redshifts less than $\sim$12,500 kms$^{-1}$.
HST WFPC2 observations then reveal that CFC97 Cen 05 is not
a dwarf member of the Cen~A group at all, rather it is a distant almost 
face-on spiral galaxy.  While mistaken identifications of this type
are presumably not common, it is a salutory lesson that group membership
of apparently low luminosity small galaxies generally needs confirmation 
via a number of techniques in order to be certain of group membership. 

\acknowledgements
This research has been supported in part by funds from the Australian
Research Council through Discovery Project grant DP0343156.
The Digitized Sky Survey was produced at the Space Telescope Science Institute
under U.S. Government grant NAG W-2166. The images of these surveys are based 
on photographic data obtained using the Oschin Schmidt Telescope on Palomar 
Mountain and the UK Schmidt Telescope. The plates were processed into the 
present compressed digital form with the permission of these institutions.

\clearpage

\end{document}